\documentclass[aps,prl,floats,twocolumn,showpacs,superscriptaddress]{revtex4}

\usepackage{natbib}
\usepackage[utf8]{inputenc}

\usepackage{amsmath, amsthm, amssymb}
\usepackage{enumitem}
\usepackage{graphics,graphicx}
\usepackage{epsfig}
\usepackage{times}
\usepackage{dcolumn}
\usepackage{url}
\usepackage{color}

\begin{document}

\renewcommand*\thesection{\arabic{section}}

\title{Modeling self-sustained activity cascades in socio-technical networks}

\author{Pablo Piedrah\'ita}
\affiliation{Instituto de Biocomputaci\'on y F\'\i sica de Sistemas 
Complejos (BIFI), Universidad de Zaragoza, Mariano Esquillor s/n, 50018 Zaragoza, Spain}

\author{Javier Borge-Holthoefer}
\email{borge.holthoefer@gmail.com}
\affiliation{Instituto de Biocomputaci\'on y F\'\i sica de Sistemas 
Complejos (BIFI), Universidad de Zaragoza, Mariano Esquillor s/n, 50018 Zaragoza, Spain}

\author{Yamir Moreno}
\affiliation{Instituto de Biocomputaci\'on y F\'\i sica de Sistemas 
Complejos (BIFI), Universidad de Zaragoza, Mariano Esquillor s/n, 50018 Zaragoza, Spain}
\affiliation{Departamento de F\'{\i}sica Te\'orica, Universidad de Zaragoza, 50009 Zaragoza, Spain}

\author{Alex Arenas}
\affiliation{Instituto de Biocomputaci\'on y F\'\i sica de Sistemas 
Complejos (BIFI), Universidad de Zaragoza, Mariano Esquillor s/n, 50018 Zaragoza, Spain}
\affiliation{Departament d'Enginyeria Inform\`atica i Matem\`atiques,
Universitat Rovira i Virgili, 43007 Tarragona, Spain}
\affiliation{IPHES, Institut Catal\`a de Paleoecologia Humana i Evoluci\'o Social, C/Escorxador s/n, 43003 Tarragona, Spain}

\begin{abstract}
The ability to understand and eventually predict the emergence of information and activation cascades in social networks is core to complex socio-technical systems research. However, the complexity of social interactions makes this a challenging enterprise. Previous works on cascade models assume that the emergence of this collective phenomenon is related to the activity observed in the local neighborhood of individuals, but do not consider what determines the willingness to spread information in a time-varying process. Here we present a mechanistic model that accounts for the temporal evolution of the individual state in a simplified setup. We model the activity of the individuals as a complex network of interacting integrate-and-fire oscillators. The model reproduces the statistical characteristics of the cascades in real systems, and provides a framework to study time-evolution of cascades in a state-dependent activity scenario.
\end{abstract}


\maketitle

The proliferation of social networking tools --and the massive amounts of data associated to them-- has evidenced that modeling social phenomena demands a complex, dynamic perspective. Physical approaches to social modeling are contributing to this transition from the traditional paradigm (scarce data and/or purely analytical models) towards a data-driven new discipline \cite{watts2004new,lazer2009life,contemanifesto,giles2012computational}. This shift is also changing the way in which we can analyze social contagion and its most interesting consequence: the emergence of information cascades. Theoretical approaches, like epidemic and rumor dynamics \cite{rapoport53a,daley1964epidemics,goffman1964generalization}, reduce these events to physically plausible mechanisms. These idealizations deliver analytically tractable models, but they attain only a qualitative resemblance to empirical results \cite{borge13cascades}, for instance regarding avalanche size distributions. Yet, the challenge of having mechanistic models that include more essential factors, like the propensity of individuals to retransmit information, still remain open.

With the availability of massive amounts of microblogging data logs, like Twitter, we are in a position to scrutinize the patterns of real activity and model them. The vast majority of models to this end are based on a dynamical process that determines individuals' activity (transmission of information), and this activity is propagated according to certain rules usually based on the idea of social reinforcement, i.e. the more active neighbors an individual has, the larger his probability to become also active, and thus to contribute to the transmission of information.   

Along these lines, the often used threshold model \cite{granovetter1978threshold} (and its networked version \cite{watts2002simple}) mimics social dynamics, where the pressure to engage a behavior increases as more friends adopt that same behavior. Briefly, the networked threshold model assigns a fixed threshold $\tau$, drawn from a distribution $0 \le f(\tau) \le 1$, to each node (individual) in a complex network of size $N$ and an arbitrary degree distribution $p_{k}$. Each node is marked as \textit{inactive} except an initial seeding fraction of active nodes, typically $\rho_{0}=1/N$. A node $i$ with degree $k_{i}$ updates its state becoming active whenever the fraction of active neighbors $a_{i}/k_{i} > \tau_{i}$. The simulation of this mechanistic process evolves following this rule until an equilibrium is reached, i.e., no more updates occur. Given this setup, the \textit{cascade condition} in degree-uncorrelated networks can be derived from the growth of the initial fraction of active nodes, who on their turn might induce the one-step-to-activation (vulnerable) nodes. 
Therefore, large cascades can only occur if the average cluster size of vulnerable nodes diverges. Using a generating function approach, this condition is met at

\begin{equation}
G''_{0}(1) = \sum_{k}k(k-1)\rho_{k}p_{k} = \langle k \rangle
\label{G1}
\end{equation}
where $\rho_{k}p_{k}$ is the fraction of nodes of degree $k$ close to their activation threshold and $\langle k \rangle$ is the average degree \cite{watts2002simple}.

For $G''_{0}(1) < \langle k \rangle$ all the clusters of vulnerable nodes are small, and the initial seed can not spread beyond isolated groups of early adopters; on the contrary, if $G''_{0}(1) > \langle k \rangle$ then small fraction of disseminators may unleash --with finite probability-- large cascades. More recently, the cascade condition has been analytically determined for different initial conditions \cite{gleeson2007seed} as well as for modular and correlated networks, while placing the threshold model in the more general context of critical phenomena and percolation theory \cite{galstyan2007cascading,gleeson2008cascades,hackett2011cascades}.

These efforts, however, have a limited scope since they can account only for one-shot events, for instance the spread of chain letters, the diffusion of a single rumor or the adoption of an innovation. In other cases, instead, empirical evidence suggests that once an agent becomes active that behavior will be sustained, and reinforced, over time \cite{borge2011structural}. This creates a form of enduring activation that will be affected and affect other agents over time in a recursive way. Indeed, cascades evolve in time \cite{borge2012locating}, as a consequence of dynamical changes in the states of agents as dynamics progress. Cascades are then events that brew over time in a system that holds some memory of past interactions. Moreover, the propensity to be active in the propagation of information sometimes depends on other factors than raw social influence, e.g., mood, personal implication, opinion, etc.

In this paper, we present a threshold model, with self-sustained activity, where system-wide events emerge as microscopical conditions become increasingly correlated. We capitalize on the classical integrate-and-fire oscillator (IFO) model by Mirollo and Strogatz \cite{mirollo1990synchronization}. In this model, each node in a network of size $N$ is characterized by a voltage-like state $m$ of an oscillator, which monotonically increases with phase $\phi$ until it reaches a fixed threshold, and then it {\em fires} (emits information to its coupled neighbors, and resets its state to 0). The pulsatile dynamics, makes that each time a node fires, the state of its $k$ neighbors is increased by $\varepsilon$. More precisely, $m \in [0,1]$ is uniformly distributed at $t=0$ and evolves such that

\begin{equation}
m = f(\phi) = \frac{1}{w}\ln(1 + [e^{w}-1] \phi)
\end{equation}
parametrized by $w > 0$ to guarantee that $f$ is concave down. Setting the fixed threshold to 1, whenever $m_{i} = 1$ then \textit{instantaneously} $m_{j} = \min(1, m_{j}+ \varepsilon)$, if the edge $(i,j)$ exists.

Integrate-and-fire models have been extremely useful to assess the bursty behavior and the emergence of cascades in neuronal systems represented in lattices \cite{corral95prl} and complex networks \cite{timme2002coexistence,roxin}. 
We propose to model social systems as a complex network of IFOs representing the time evolving (periodic in this case) activation of individuals. The model comprises two free parameters, $w$ and $\varepsilon$, which are closely related. Dissipation $w$ may be interpreted as the willingness or \textit{intrinsic propensity} of agents to participate in a certain diffusion event: the larger $w$, the shorter it takes for a node to enter the tip-over interval $1-\varepsilon < m < 1$. 
Conversely, $\varepsilon$ quantifies the amount of influence an agent exerts onto her neighbors when she shows some activity. Larger $\varepsilon$'s will be more consequential for agents, forcing them more rapidly into the tip-over region. 
Both quantities affect the level of \textit{motivation} $m$ of a given agent. Note that $\varepsilon$ in the current framework maps onto $\tau$ in the classical threshold model, in the sense that both determine the width of the tip-over region. Finally, the phase is translated into time steps, and then prescribed as $\phi(t)=t$. 

We use Eq.~\ref{G1} to derive the cascade condition in this new framework. Note that now the distribution of activity is governed by $\rho(t) = 1- \int_0^{1-\varepsilon} g(m,t)\,dm$ where $g(m,t)$ corresponds to the states' probability distribution at a certain time $t$. For an initial uniform distribution of motivation $m$ and a fixed $\varepsilon$, the condition for the emergence of cascades reads

\begin{equation}
G''_{0}(1)_{t=0} = \varepsilon \sum_{k}k(k-1) p_{k} = \langle k \rangle
\label{G2}
\end{equation}

and in general for any time,

\begin{equation}
G''_{0}(1)_{t} = \rho(t) \sum_{k}k(k-1) p_{k} = \langle k \rangle,
\label{G3}
\end{equation}

\noindent which implies that the cascade condition depends on time in our proposed framework. It is worth noticing that, in this scenario, $\rho(t)$ is not a function of the node degree $k$.

\begin{figure}[tbp]
\begin{center}
  \includegraphics[width=\columnwidth,clip=0]{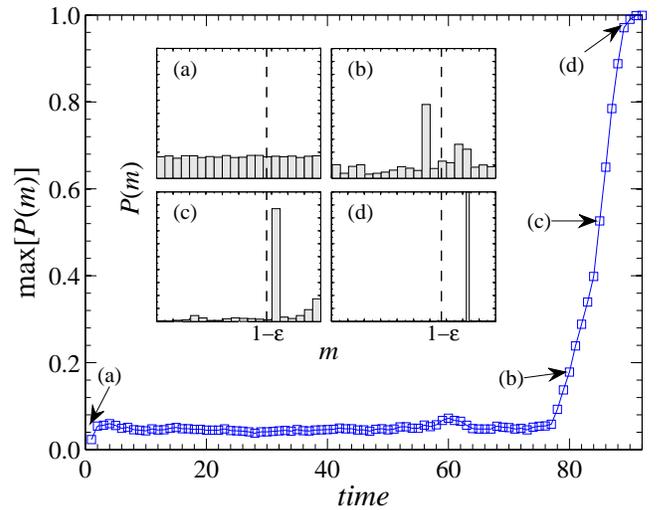}
\caption{(color online). Inset (a)-(d): Motivation $m$ probability distributions of four different representative times along the synchronization window. Each snapshot depicts the $m$-state histogram of the $N$ oscillators. The dynamics begins with a random uniform distribution of $m$-states --inset (a)-- and it progressively narrows during the transition to synchrony --inset (d). Main: largest fraction of synchronized nodes across time. The path to synchronization evolves steadily at a low level, and eventually suffers an abrupt transition.}
\label{fig1}
\end{center}
\end{figure}

As the dynamics evolve in time, the states of the nodes progressively correlate and, consequently, the distribution of states changes dramatically. The evolution of the states distribution is depicted in Fig.~\ref{fig1}. The initially uniform distribution $g(m,0)$ (inset (a)) evolves towards a Dirac $\delta$ function (inset (d)) as the network approaches global synchronization. We have not been able to find a closed analytical expression for the consecutive composition of the function $g(m,t)$ after an arbitrary number of time steps to reveal the evolution of $\rho(t)$, nonetheless it can be solved numerically. Eq.~\ref{G3} reduces to

\begin{equation}
 \rho(t)  (\langle k^{2} \rangle - \langle k \rangle)= \langle k \rangle.
\label{G4}
\end{equation}

The cascade condition is thus

\begin{equation}
\frac{\rho(t)}{1+\rho(t)} = \frac{\langle k \rangle}{\langle k^{2} \rangle},
\label{G5}
\end{equation}

\noindent that exactly corresponds to the bond percolation critical point on uncorrelated networks \cite{molloy1995critical,molloy1998size,dorogovtsev08}.
For the case of random Poisson networks $\langle k^{2} \rangle \sim \langle k \rangle^{2}$, then 
\begin{equation}
\frac{\rho(t)}{1+\rho(t)} = \frac{1}{\langle k \rangle}
\label{G6}
\end{equation}

\begin{figure}[tbp]
\begin{center}
  \includegraphics[width=\columnwidth,clip=0]{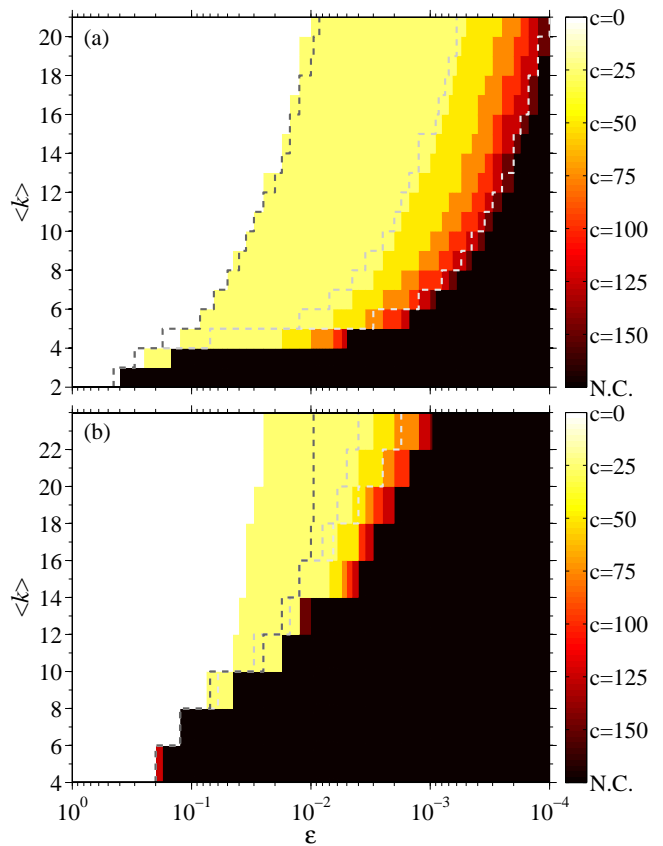}
\caption{(color online)$(\varepsilon, \langle k \rangle)$ cascade diagram for different cycles (coded by color), with fixed $w=3$. Vertical axis and each dashed line define a confined region in which global cascades might occur according to Eq.~\ref{G5} and for a specific cycle (here we show only the expected zones for $c=0$ --dashed white-- and $c=150$ --dashed gray). Results are obtained for synthetic Erd\"{o}s-R\'enyi (a) and scale-free with $\gamma= 3$ (b) uncorrelated networks of size $N=10^{4}$. A cascade is considered ``macroscopical'' if the synchronized cluster $S_{c} \ge 0.25N$. Color codes indicate the existence of at least one cascade $S > S_{c}$ in numerical simulations; analytical predictions are averaged over 200 networks with random initial conditions. Note that the cascade condition in (a) often underestimates the actual cascade regions because it does not take into account second order interactions; the same applies in the lower panel (b), except for $c=0$ where the analytical prediction overestimates the results because the inclusion of the hub into the cascade is improbable starting from a uniform distribution.} 
\label{fig2}
\end{center}
\end{figure}

We can now explore the cascade condition in the $(\varepsilon, \langle k \rangle)$ phase diagram in Fig.~\ref{fig2}, and compare the analytical predictions with results from extensive numerical simulations. Since the time to full synchronization (global cascade) is different for each $(\varepsilon, \langle k \rangle)$, we introduce \textit{cycles}. One cycle is complete whenever every node in the network has fired at least once. In this way we bring different time scales to a common, coarse-grained temporal ground, allowing for comparison. The regions where cascades above a prescribed threshold $S_{c}$ are possible are color-coded for each cycle ${0, 25, 75, 100,\ldots}$, and black is used in regions where cascades do not reach $S_{c}$ (labeled as N.C., ``no cascades'', in Fig.~\ref{fig2}). Note that if cascades are possible for a cycle $c$, they will be possible also for any $c' \ge c$. This figure renders an interesting scenario: on the one hand, it suggests the existence of critical $\varepsilon_{c}$ values below which the cascade condition is systematically frustrated (black area in the phase diagram). On the other, it establishes how many cycles it takes for a particular $(\varepsilon, \langle k \rangle)$ pair to attain macroscopical cascades (full synchronization) --which becomes an attractor thereafter, for undirected connected networks. Given the cumulative dynamics of the current framework, in contrast with Watts' model, the region in which global cascades are possible grows with $\langle k \rangle$.

Turning to the social sphere, these results open the door to predicting how long it takes for a given topology, and a certain level of inter-personal influence, to achieve system-wide events. Furthermore, the existence of a limiting $\varepsilon_{c}$ determines whether such events can happen at all.

Additionally, the predictions resulting from Eq.~\ref{G5} are represented as dashed lines in Fig.~\ref{fig2}. For the sake of clarity, we only include predictions for $c=0$ (dashed black), $c=25$ (dashed gray) and $c=150$ (dashed white). Projections from this equation run close to numerical results in both homogeneous (Fig.~\ref{fig2}a) and inhomogeneous networks (Fig.~\ref{fig2}b), although some deviations exist. Noteworthy, Eq.~\ref{G5} clearly overestimates the existence of macroscopic cascades in the case of scale-free networks at $c=0$. Indeed, $\rho(0) = \varepsilon$ does not yet incorporate the inherent dynamical heterogeneity of a scale-free topology, thus Eq.~\ref{G5} is a better predictor as the dynamics loose memory of the hardwired initial conditions. In the general case $c > 0$, deviations are due to the fact that the analytical approach in the current work is not developed beyond first order. Second order corrections to this dynamics (including dynamical correlations) should be incorporated to the analysis in a similar way to that in \cite{gleeson2008cascades}, however it is beyond the scope of the current presentation.

\begin{figure}[tbp]
\begin{center}
  \includegraphics[width=\columnwidth,clip=0]{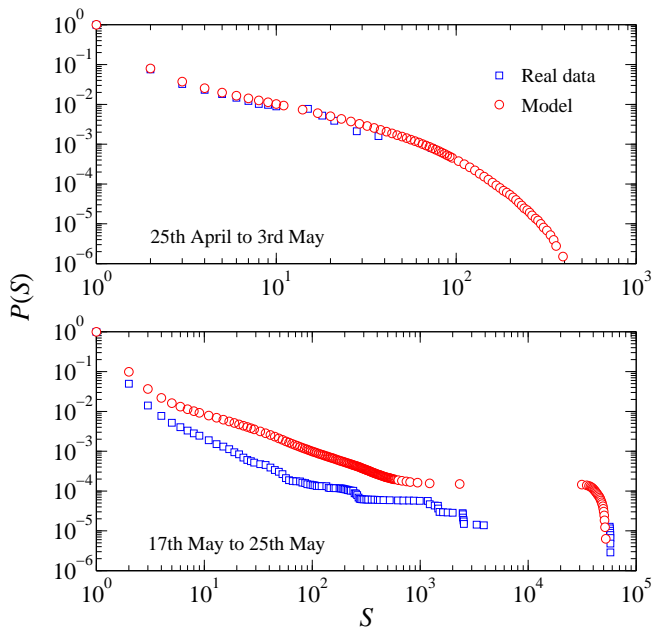}
\caption{(color online). Cascade size cumulative distributions of real data (blue squares) and the model counterpart (red circles). We have considered two time windows which significantly differ: first eight days (top) for which we have set $w=0.1$ and $\varepsilon = 0.0020 < \varepsilon_{c}$; last eight days (bottom) for which we have  $w=30.0$ and $\varepsilon = 0.0012 > \varepsilon_{c}$. Note that $\varepsilon_{c}$ varies for different $w$ values. The model performs well in both periods, the relative error of the slope in the linear region is $< 2\%$. Real data distributions are measured as in \cite{gonzalez2011dynamics,borge2012locating}.}
\label{fig3}
\end{center}
\end{figure}

According to Mirollo \& Strogatz \cite{mirollo1990synchronization}, synchronicity emerges more rapidly when $w$ or $\varepsilon$ is large; then the time taken to synchronize is inversely proportional to the product $w\varepsilon$. In our simulations, we use this cooperative effect between coupling and willingness to fix $\varepsilon$, which is set to a set of values (slightly above or below) $\varepsilon \simeq \varepsilon_{c}$, and use $w$ to fine-tune the matching between observed cascade distributions and our synthetic results (see Fig. \ref{fig3}).

To illustrate the explanatory power of the dynamical threshold model, we use data from \url{www.twitter.com}. They comprise a set of $\sim0.5$ million Spanish messages publicly exchanged through this platform from the 25th of April to the 25th of May, 2011 \cite{15Mdata}. In this period a sequence of civil protests and demonstrations took place, including camping events in the main squares of several cities beginning on the 15th of May and growing in the following days. Notably, a pulse-based model suits well with the affordances of this social network, in which any emitted message is instantly broadcasted to the author's immediate neighborhood --its set of \textit{followers}. For the whole sample, we queried for the list of followers for each of the emitting users, discarding those who did not show outgoing activity during the period under consideration. The set of users $N=87569$ plus their following relations constitute the topological support (directed network) for the dynamical process running on top of it. The average degree of this network is $\langle k \rangle = 69$ and its degree distribution scales like $p(k) = k^{-1.5}$.

On top of the described network, we measure the cascade size distribution for different periods as in \cite{borge2012locating,gonzalez2011dynamics}. In Fig.~\ref{fig3} these periods correspond to the ``slow-growth'' phase (25th April to 3rd May; blue squares in the upper panel) and to the ``explosive'' phase (19th to 25th May; blue squares in the lower panel), which comprehends the most active interval --the reaction to the Spanish government ban on demonstrations around local elections on the 22nd May. On the other hand, we run the proposed dynamics on the same topology for different $w$ values, with remarkable success (red circles). Interestingly, an appropriate fitting is attained when $w$ is adapted to real-world excitation level: when cascades are measured in an interval around the 15th May, a higher $w$ is needed --nominating it as a suitable proxy for the system's excitability. Additionally, modeling low-activity periods can be achieved just by setting $\varepsilon < \varepsilon_{c}$.

Summarizing, we have proposed a time-dependent continuous self-sustained model of social activity. The model can be analyzed in the context of previous cascade models, and encompasses new phenomenology as the time-dependence of the critical value of the emergence of cascades. We interpret it under a social perspective, where collective behavior is seen as an evolving phenomenon resulting from inter-personal influence, contagion and memory. In a general perspective, our modeling framework offers an alternative approach to the analysis of interdependent decision making and social influence. It complements threshold models and complex contagion taking into account time dynamics and recursive activation, and also splits motivation into two components: intrinsic propensity and strength of social influence. We also anticipate that the exploration of the whole parametric space would lead to new insights about the effects of social influence and interdependence in social collective phenomena.
 
\section{Acknowledgements}
We gratefully acknowledge Sandra Gonz\'alez-Bail\'on for interesting discussions and comments on the draft. This work has been partially supported by MINECO through Grant FIS2011-25167 and FIS2012-38266; Comunidad de Arag\'on (Spain) through a grant to the group FENOL and by the EC FET-Proactive Project PLEXMATH (grant 317614). A. A. also acknowledges partial financial support from the ICREA Academia and the James S. McDonnell Foundation. 


\end{document}